\documentclass{INTERSPEECH2023}
\usepackage{amsmath,graphicx}
\usepackage{amssymb,xcolor,url,cite}
\usepackage{booktabs,comment,bm}
\usepackage{here}
\usepackage{multirow}
\usepackage[acronym,shortcuts]{glossaries}
\interspeechcameraready

\title{Multi-Stream Extension of Variational Bayesian HMM Clustering (MS-VBx) \\for Combined End-to-End and Vector Clustering-based Diarization}

\name{
\begin{tabular}{c}
Marc Delcroix$^1$, Naohiro Tawara$^1$, Mireia Diez$^2$, Federico Landini$^2$, Anna Silnova$^2$, \\ \textit{Atsunori Ogawa}$^1$, \textit{Tomohiro Nakatani}$^1$,\textit{Lukas Burget}$^2$, \textit{Shoko Araki}$^1$
\end{tabular}
}
\address{{$^1$}NTT Corporation, Japan, {$^2$}Brno University of Technology, Speech@FIT}

\email{marc.delcroix@ieee.org}

\def\S{{\mathbf{S}}}
\def\S[#1]{{\mathbf{S}_{#1}}}

\def\X{{\mathbf{X}}}
\def\x[#1]{{\mathbf{x}_{#1}}}
\def\xt{{\mathbf{x}_t}}
\def\xti[#1]{{\mathbf{x}^{#1}_t}}

\def\a[#1]{{\mathbf{a}_{#1}}}
\def\at{{\mathbf{a}_t}}
\def\ati[#1]{{\mathbf{a}^{#1}_t}}

\def\W{{\mathbf{W}}}
\def\wt{{w_t}}
\def\wi[#1]{{w_{#1}}}

\def\Z{{\mathbf{Z}}}
\def\z[#1]{{z_{#1}}}
\def\zt{{\z[t]}}

\def\Y{{\mathbf{Y}}}

\def\ysi[#1]{{\mathbf{y}_{s,{#1}}}}
\def\y[#1]{{\mathbf{y}_{#1}}}

\def\I{{\mathbf{I}}}

\def\alphas{{\bm{\alpha}_s}}
\def\alphas[#1]{{\bm{\alpha}_{{#1}}}}

\def\alphasi[#1]{{\bm{\alpha}_{s,{#1}}}}
\def\alphasiT[#1]{{\bm{\alpha}_{s,{#1}}^{\top}}}

\def\Lsi[#1]{{\mathbf{L}_{s,{#1}}}}
\def\Lsiinv[#1]{{\mathbf{L}^{-1}_{s,{#1}}}}
\def\Lsinv[#1]{{\mathbf{L}^{-1}_{{#1}}}}

\def\gammats{{\gamma_{t,s}}}
\def\gammatsi[#1]{{\gamma_{t,s,#1}}}

\def\yg{{\mathbf{y}_g}}

\def\Lg{{\mathbf{L}_g}}
\def\Lginv{{\mathbf{L}_g^{-1}}}
\def\alphag{{\bm{\alpha}_g}}

\def\rhotc{{\bm{\rho}^c_t}}
\def\psimat{{\bm{\Phi}}}

\def\V{{\mathbf{V}}}

\newacronym{AHC}{AHC}{agglomerative hierarchical clustering}
\newacronym{cAHC}{cAHC}{constrained agglomerative hierarchical clustering}
\newacronym{DER}{DER}{Diarization error rate}
\newacronym{EEND}{EEND}{end-to-end diarization}
\newacronym{EEND-VC}{EEND-VC}{\gls{EEND} combined with \gls{VC}}
\newacronym{EEND-GLA}{EEND-GLA}{\gls{EEND} with
global and local attractors}
\newacronym{ELBO}{ELBO}{evidence lower bound objective}
\newacronym{GMM}{GMM}{Gaussian mixture model}
\newacronym{HMM}{HMM}{hidden Markov model}
\newacronym{iGMM}{iGMM}{infinite GMM}
\newacronym{LDA}{LDA}{linear discriminant analysis}
\newacronym{MS-VBx}{MS-VBx}{multi-stream \gls{VBx}}
\newacronym{OSD}{OSD}{overlapped speech detection}
\newacronym{PLDA}{PLDA}{probabilistic linear discriminant analysis}
\newacronym{TS-VAD}{TS-VAD}{target speaker voice activity detection}
\newacronym{SAD}{SAD}{speech activity detection}
\newacronym{VBx}{VBx}{variational Bayesian HMM clustering of x-vector sequences}
\newacronym{VB}{VB}{variational Bayes}
\newacronym{VB-GMM}{VB-GMM}{\gls{VB}-Gaussian mixture model}
\newacronym{VC}{VC}{vector clustering}

\begin{document}
\setlength{\abovedisplayskip}{5pt}
\setlength{\belowdisplayskip}{3pt}

\maketitle
 \begin{abstract}
Combining end-to-end neural speaker diarization (EEND) with vector clustering (VC), known as EEND-VC, has gained interest for leveraging the strengths of both methods. EEND-VC estimates activities and speaker embeddings for all speakers within an audio chunk and uses VC to associate these activities with speaker identities across different chunks. EEND-VC generates thus multiple streams of embeddings, one for each speaker in a chunk. We can cluster these embeddings  using constrained agglomerative hierarchical clustering (cAHC), ensuring embeddings from the same chunk belong to different clusters. This paper introduces an alternative clustering approach, a multi-stream extension of the successful Bayesian HMM clustering of x-vectors (VBx), called MS-VBx. Experiments on three datasets demonstrate that MS-VBx outperforms cAHC in diarization and speaker counting performance.
\end{abstract}

\noindent\textbf{Index Terms}: speaker diarization, end-to-end, VBx, clustering

\section{Introduction}
\label{sec:intro}

Diarization consists of determining who speaks when in a multi-talker recording. It plays an essential role in the processing of conversations. 
There are several approaches to tackle the diarization problem, such as \emph{\gls{VC}}\cite{DIHARD_JHU}, \emph{\gls{EEND}}~\cite{Fujita_IS2019, Horiguchi2022_EDA_EEND} and \emph{\gls{TS-VAD}}~\cite{Medennikov2020}. Recently, the combination of the first two, i.e., \gls{EEND-VC} has received increased interest since it offers a principled way of getting the best of both frameworks achieving high performance on several tasks~\cite{EEND-vector-clustering_ICASSP2021, Horiguchi2021_EEND_offline_GLA, WavLM}.
The term \gls{EEND-VC} was introduced in~\cite{EEND-vector-clustering_ICASSP2021}, but here it refers to a larger class of approaches that combine \gls{EEND} and \gls{VC} including, e.g.,~\cite{Horiguchi2021_EEND_offline_GLA}. 
This paper discusses a novel clustering approach for \gls{EEND-VC}.

\gls{VC}-based diarization approaches first compute speaker embedding vectors for short segments of a recording and then cluster these embeddings to assign speaker labels to each segment. 
Classical clustering algorithms such as K-means or \gls{AHC} can be used, but more powerful clustering schemes have been proposed, such as \gls{VBx}~\cite{diez19_interspeech, landini2022bayesian}, which has been widely used in diarization challenges~\cite{landini2020but, wangustc_DIHARDIII, yu2022summary}. \gls{VC} approaches can work with an arbitrarily large number of speakers in a recording. However, they assume no speech overlap in a segment, which does not hold for many natural conversations. 

\gls{EEND}~\cite{Fujita_IS2019, Fujita_ASRU2019, Horiguchi2022_EDA_EEND} is an alternative approach, which uses a neural network to directly output the speech activity for each speaker in a recording even for overlapping regions. \gls{EEND} can thus handle overlapping speech, making it a competitive alternative to \gls{VC}.
However, it is more challenging to generalize to an arbitrarily large number of speakers~\cite{Horiguchi2022_EDA_EEND}. 

\begin{figure}[tb]
 \centering
 \includegraphics[width=0.43\textwidth]{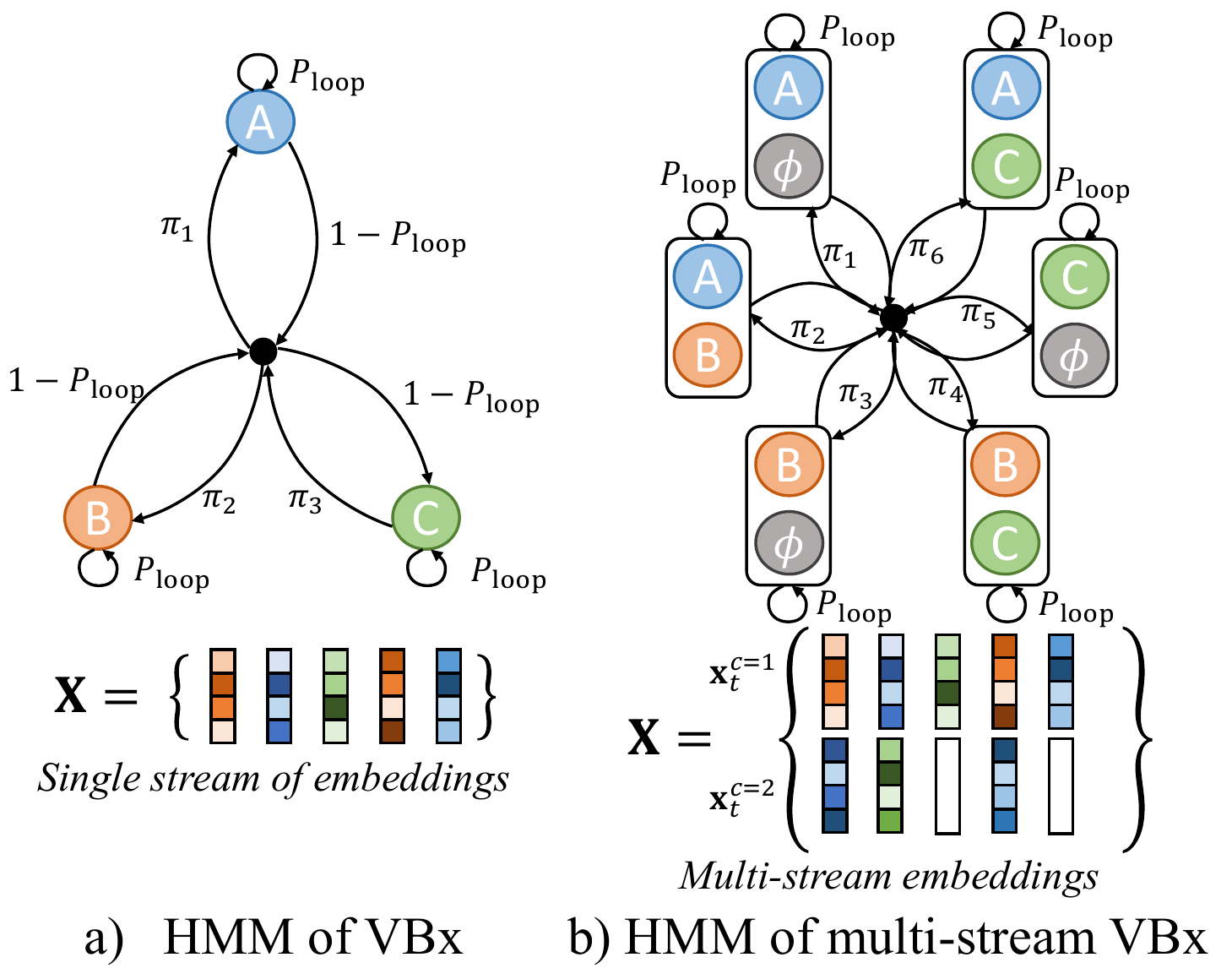}
 \vspace{-2mm}
 \caption{Schematic diagram of the \gls{HMM} used for a) the conventional \gls{VBx} and b) the proposed multi-stream extension ($C=2$). ``A,'' ``B,'' and ``C'' represent speakers and $\phi$ an inactive speaker. $P_{loop}$ is the loop probability. For simplicity, we omitted the states ``B A,'' ``C A,'' ``C B''.}
 \label{fig:vbx}
 \vspace{-6mm}
\end{figure}

\gls{EEND-VC} has been proposed to combine the strength of both frameworks~\cite{EEND-vector-clustering_ICASSP2021,Horiguchi2021_EEND_offline_GLA}. It first performs \gls{EEND} on speech \emph{chunks}\footnote{Typically, the chunks of \gls{EEND-VC} are speech segments longer than those used by \gls{VC}.} using a modified network architecture~\cite{EEND-vector-clustering_ICASSP2021, Horiguchi2021_EEND_offline_GLA}, which estimates the speaker activities and speaker embeddings. Then, it performs \gls{VC} on the estimated speaker embeddings to stitch together the speaker activities of the same speaker across different chunks.
\gls{EEND-VC} can thus handle overlapping speech as \gls{EEND} and an arbitrary number of speakers as \gls{VC}.

Compared to conventional \gls{VC}, with \gls{EEND-VC}, each chunk can have multiple speaker embeddings. Therefore, it requires \emph{clustering multi-stream of speaker embeddings} with a constraint that the embeddings from the same chunk should not belong to the same cluster~\cite{EEND-vector-clustering_Interspeech2021}. 
This has been ensured by using constrained variants of well-known clustering algorithms, such as constrained variants of K-means~\cite{COP-kmeans, Yang2013_CLC_Kmeans} or \gls{cAHC}~\cite{constrained_AHC}. 
However, variational Bayesian HMM clustering of x-vector sequences (VBx)~\cite{diez19_interspeech,landini2022bayesian} has shown better performance than K-means or AHC in diarization and has been widely used in recent challenges~\cite{landini2020but, wangustc_DIHARDIII, yu2022summary}.
 In this paper, we investigate the use of \gls{VBx} for clustering the speaker embeddings of \gls{EEND-VC} by developing a multi-stream extension of \gls{VBx}.

\gls{VBx} is a clustering algorithm based on a Bayesian HMM where states are associated with the speakers and transitions between states represent speaker turns, as shown in Fig. \ref{fig:vbx}-a). \gls{VBx} models the speaker distributions with Gaussians and sets a prior on the parameters of the Gaussians derived from a \gls{PLDA} model trained on a large set of speaker embeddings. 
The parameters of the \gls{HMM}, the speaker models, and the best assignment of states given a sequence of embeddings are estimated directly on each recording using the \gls{VB} inference.
Compared to other clustering approaches, \gls{VBx} allows modeling the transitions between speakers and offers a principled way of estimating the number of speakers through the \gls{VB} inference~\cite{Diez_taslp_2020}.
The formulation of \gls{VBx} assumes a single speaker for each segment and thus for each embedding. 

We propose to generalize the \gls{VBx} algorithm to handle the multi-stream embeddings of \gls{EEND-VC}, where each \gls{HMM} state corresponds to a single or multiple (overlapping) speakers as shown in Fig. \ref{fig:vbx}-b). We call this extension \gls{MS-VBx}. 
A naive implementation would exponentially increase the number of parameters of the model and also make the allocation of the speakers difficult. Therefore, we propose to tie together parameters of the \gls{HMM} states that correspond to the same speaker. This generalization of \gls{VBx} allows using the state-of-the-art \gls{VC} approach for \gls{EEND-VC}.

We show that the proposed \gls{MS-VBx} naturally implements the constraint required by \gls{EEND-VC}. Besides, it outperforms \gls{cAHC} in experiments on the CALLHOME and DIHARD II and III datasets in terms of \gls{DER} and speaker counting errors.

\section{Overview of EEND-VC}
\begin{figure}[tb]
 \centering
 \includegraphics[width=0.4\textwidth]{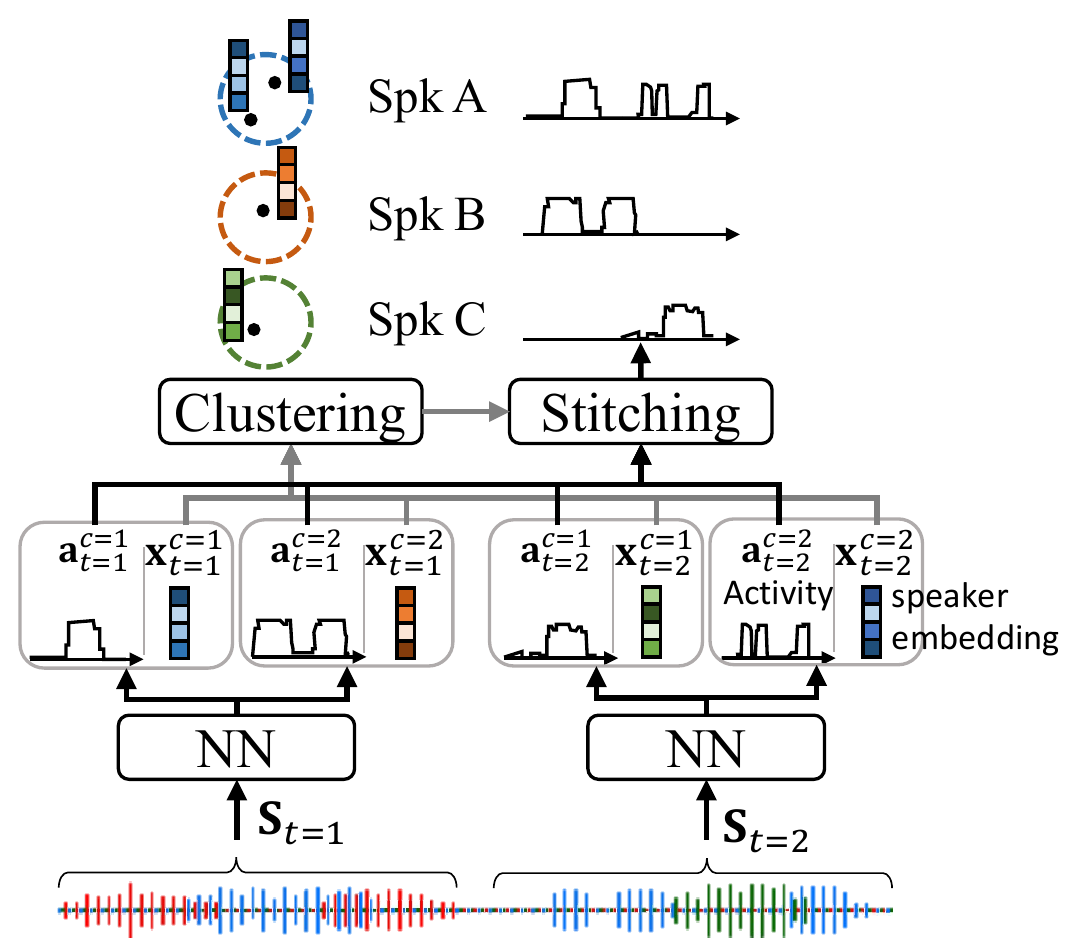}
 \vspace{-2mm}
 \caption{Schematic diagram of \gls{EEND-VC}.}
 \label{fig:eendvc}
 \vspace{-6mm}
\end{figure}

Figure \ref{fig:eendvc} shows an \gls{EEND-VC} system.
$\S[t]$ are the speech features for the chunk $t=1, \ldots, T$.
\gls{EEND-VC} consists of a neural network that estimates for each chunk $t$, the speech activities $\at = [\ati[1], \ldots, \ati[C] ] \in [0,1]^{N\times C}$ and speaker embeddings $\xt=[\xti[1], \ldots, \xti[C]] \in \mathbb{R}^{D \times C}$. Here, $N$ represents the number of time frames per chunk, $D$ the dimension of the speaker embedding vectors, and $C$ the number of output streams, which can be a fixed maximum number of active speakers in a chunk~\cite{EEND-vector-clustering_ICASSP2021} or an estimate of the number of active speakers for each chunk~\cite{Horiguchi2021_EEND_offline_GLA}.
$\ati[c]$ represents the speech activity of the speaker associated with the output stream $c$, and $\xti[c]$ is the speaker embedding associated with that speaker. 

\gls{EEND-VC} processes each chunk independently. Since the total number of speakers in a recording can exceed the number of speakers in a chunk, and the system can output speakers in an arbitrary order, there is speaker ambiguity at the output of the \gls{EEND} stage. \gls{EEND-VC} resolves this ambiguity by clustering the speaker embeddings to associate a global speaker identity to each estimated speech activity, $\ati[c]$. We can then produce the diarization results by stitching the speech activities associated with the same speaker identity.

Prior works have explored various clustering approaches and reported superior performance with \gls{cAHC}~\cite{EEND-vector-clustering_Interspeech2021}. In this paper, we propose an alternative clustering approach. 

\section{Proposed multi-stream VBx}

As mentioned in section \ref{sec:intro}, the original derivation of \gls{VBx} assumes a single embedding vector per segment and no overlap. Consequently, we need to generalize it if we want to use it for clustering the multi-stream embeddings of \gls{EEND-VC}.

\subsection{Multi-stream extension of VBx model}
 \gls{MS-VBx} generalizes the \gls{VBx} algorithm to multi-stream embeddings. With \gls{MS-VBx}, multiple speakers are associated with an \gls{HMM} state, and the parameters corresponding to the same speakers are tied together across \gls{HMM} states, as shown in Figure \ref{fig:vbx}-b). 
 We design the \gls{HMM} such that the same speaker cannot appear more than once in a given state, i.e., there are no states such as ``A A,'' ``B B,'' or ``C C.'' This design naturally implements the cannot-link constraint required by \gls{EEND-VC}. 
 In the following, we denote the \gls{HMM} state index by $s=1,\ldots, S$, and the speaker index by $g=1, \ldots, S_g$.

Similarly to VBx, \gls{MS-VBx} aims at finding the most likely sequence of latent variables $\Z = \{\z[1], \ldots, \z[T] \}$, where $\z[t]$ defines the hard alignment of the embedding vectors to the HMM states for chunk $t$. 
The complete model is expressed as,
\begin{align}
 p(\X, \Z, \Y) & = p(\X|\Z, \Y) p(\Z) p(\Y) \label{eq:full_prob} \\
   & = \prod_t p(\xt|\zt) p(\zt|\z[t-1]) \prod_g p(\yg),\label{eq:full_prob2}
\end{align} 
where $\X = \{\x[1], \ldots, \x[T]\}$ is the sequence of speaker embeddings, $\Y =\{\y[1], \ldots, \y[S_g]\}$ is the set of all the speaker-specific latent variables, $\y[g]$, and $p(\zt|\z[t-1])$ represents the state transition probability. 

The output probability of \gls{HMM} state $s$, $p(\xt|\zt)$, represents the probability that speakers associated with that state are active in that chunk.\footnote{We assume here for simplicity that there is an active speaker for each stream, $c$, unlike what Fig \ref{fig:vbx}-b) suggests. We deal with the possibility of having inactive speakers in Section \ref{ssec:inactive speakers}.} It is given by,
\begin{align}
 p(\xt|\zt = s) = & \prod_{c=1}^C p(\xti[c]|\zt = s),
\end{align}
where $p(\xti[c]|\zt = s) = \mathcal{N}(\xti[c]; \V \y[\tilde{g}], \I)$,
and $p(\y[\tilde{g}]) = \mathcal{N}(\y[\tilde{g}];\mathbf{0}, \I)$. Here, $\tilde{g}$ is given by $\tilde{g}=\text{Spk}(s,c)$, where $\text{Spk}(s,c)$ is a mapping function that maps the state and stream indexes $(s,c)$ to the speaker index, $\tilde{g}$. In other words, we tie together the Gaussian parameters associated with the same speaker across HMM states and denote by $\mathcal{S}_g$ the set of HMM-sub-states $(s,c)$ associated with the speaker, $g$. We then have $\text{Spk}(s,c) = g, \forall (s,c) \in \mathcal{S}_g$.

Following the original \gls{VBx}~\cite{landini2022bayesian}, $\yg$ is a latent variable acting as a prior on the mean of the speaker models $p(\xti[c]|\zt = s)$, which is derived from a pre-trained \gls{PLDA} model. $\V = \psimat^{\frac{1}{2}}$ is a feature transformation matrix, and $\psimat$ is a diagonal matrix corresponding to the between-speaker covariance matrix in the transformed space of the \gls{PLDA} model.

As in the original \gls{VBx}, we set the transition probability to $p(\zt=s|\z[t-1]=s') = (1-P_{\text{loop}}) \pi_s + \delta_{s,s'} P_{\text{loop}}$, where $P_{\text{loop}}$ is the loop probability, $\pi_s$ is the probability to transition to state $s$ from the non-emitting node, and $\delta$ is the Kronecker delta. $P_{\text{loop}}$ is a tuning parameter. We estimate $\pi_s$ with Bayesian inference.

\subsection{Inference}

The most likely sequence $\Z$ is obtained using \gls{VB} inference.
This is solved by iteratively updating the approximate posterior distribution $q(\Y)$ for a fixed approximate posterior $q(\Z)$ and vice versa.
The distribution $q(\Y)=\prod_g q(\yg)$ is updated as, 
\begin{align}
 q(\yg) & = \mathcal{N}(\mathbf{y}; \alphag, \Lginv), 
\end{align}
where
\begin{align}
 \Lg 
 & = \I + \frac{F_A}{F_B} \biggl( \sum_t \sum_{s\in \mathcal{S}_g} \gammats \biggr) \psimat, \label{eq:var}\\
 \alphag & = \frac{F_A}{F_B} \Lginv \sum_t \sum_{(s,c)\in \mathcal{S}_g} \gammats \rhotc, \label{eq:mean}
\end{align}
 and $\rhotc= \V^{\top}\xti[c]$. $F_A$ and $F_B$ are scaling factors for the \gls{ELBO}~\cite{Diez_taslp_2020}. 
 
As for updating $q(\Z)$, the state occupancy, $\gammats$, is computed with the forward-backward algorithm using the state output probabilities obtained as,
\begin{align}
 \log\bar{p}(\xt|s) = F_A \sum_c &\left( 
 %\alphasiT[c]\rhotc -\frac{1}{2} \text{tr} 
 %\left(\psimat (\Lsiinv[c]+ \alphasi[c]\alphasiT[c]) \right)
 \alphas[\tilde{g}]\rhotc -\frac{1}{2} \text{tr} 
 \left(\psimat (\Lsinv[\tilde{g}]+ \alphas[\tilde{g}]\alphas[\tilde{g}]) \right)
 \right.\nonumber \\
 & \left. - \frac{D}{2} \ln(2\pi) -\frac{1}{2} \xti[c]^{\top}\xti[c] \right),\label{eg:state_output}
\end{align}
where $\tilde{g} = \text{Spk(s,c)}$, and $\text{tr}(\cdot)$ is the trace operator.
Note that $\Lg$ is a diagonal matrix because $\psimat$ is diagonal, making the matrix inversion in Eqs. \eqref{eq:mean} and \eqref{eg:state_output} trivial.

We can estimate $\pi_s$ as in the conventional \gls{VBx}~\cite{landini2022bayesian}.
The inference process tends to set $\pi_s$ to zero for redundant states, dropping these states~\cite{landini2022bayesian}. With the original \gls{VBx}, the number of remaining states provides a direct estimate of the number of speakers. For \gls{MS-VBx}, we also derive the number of speakers from the remaining \gls{HMM} states, knowing the association between speakers and states. 

The inference with \gls{MS-VBx} is similar to that of the conventional \gls{VBx}, except for the use of tied Gaussians, which introduces the summation over the set of HMM-sub-states $\mathcal{S}_g$ in Eqs. \eqref{eq:var} and \eqref{eq:mean}, and the summation over the different streams in Eq. \eqref{eg:state_output}. \gls{MS-VBx} can thus be easily implemented by extending an existing \gls{VBx} code.\footnote{\url{https://github.com/BUTSpeechFIT/VBx}}

\subsection{Handling inactive speakers}
\label{ssec:inactive speakers}
\def\pnovt{p^{1}_t}
The original \gls{VBx} removes the silent portions of the signals beforehand since it is challenging to model silent segments with speaker embeddings. In our case, there are chunks where the number of active speakers may be smaller than the number of streams, $C$. We denote by $\hat{C}_t$ the estimated number of active speakers for chunk $t$, and by $C_s$ the number of active speakers for the \gls{HMM} state $s$. 

\gls{EEND-VC} can provide information about the number of active speakers in a chunk. For example, the \gls{EEND-GLA} approach~\cite{Horiguchi2021_EEND_offline_GLA} has a variable number of outputs corresponding to the number of active speakers in a chunk. In contrast, the \gls{EEND-VC} scheme~\cite{EEND-vector-clustering_Interspeech2021} has a fixed number of outputs corresponding to the maximum number of speakers in a chunk but is trained to output speech activity close to zero for outputs with no active speakers. We can thus consider that an output, $c$, is inactive if $\frac{1}{N} \sum_{n=1}^N a_{t,n}^c < \tau $, where $\tau$ is a predetermined threshold.

We reformulate Eq. \eqref{eq:full_prob2} by introducing a random variable $\W=[\wi[1], \ldots, \wi[T] ]$ representing the number of speakers in a chunk:
\begin{align}
 p(\X, \W, \Z, \Y) %= &p(\X, \V| \Z, \Y) p(\Z) p(\Y) \\
   = & \prod_t p(\xt, \wt|\zt) p(\zt|\z[t-1]) \prod_g p(\yg),
\end{align} 
where we express the state emission probabilities as,
\begin{align}
 p(\xt, \wt=C_s|\zt = s) & = p(\wt=C_s) \prod_{c=1}^{C_s} p(\xti[c]|\zt = s), \nonumber
\end{align}
and $p(\wt=C_s)=\delta_{\hat{C}_t,C_s}$. %, where $\delta$ is the Kronecker delta.
We thus rely on the hard decision from \gls{EEND-VC} to determine $\wt$. 

\subsection{Overall procedure}
We first train a \gls{PLDA} model on the speaker embeddings of the \gls{EEND-VC} model extracted from the training/adaptation datasets.
At inference time, we first compute speaker activities and embeddings with \gls{EEND-VC} for each recording. Then, as with standard \gls{VBx}, we perform \gls{cAHC} to obtain a rough estimate of the number of speakers (which \gls{VBx} can refine/reduce), generate the \gls{HMM} and initialize the state occupancy. Finally, we run \gls{MS-VBx} and stitch the activities, $\ati[c]$, based on the clustering results, to obtain the diarization results.
Note that compared to standard VBx, MS-VBx requires more
HMM states for the same total number of speakers, increasing thus the  computational complexity. 
Future works should address this issue. 

\section{Related works}
Another \gls{VB} scheme has been proposed recently for clustering of \gls{EEND-VC}~\cite{Kinoshita_tight_2022}. That work aimed to allow training speaker embedding jointly with the clustering algorithm to reduce the mismatch between training and inference. The scheme relied on using \gls{VB-GMM} such as \gls{iGMM}, which are more complex models, as they assume a potentially infinite number of speakers.
Besides, with \gls{iGMM}, it is challenging to implement the cannot-link constraint.
In this paper, we base our study on \gls{VBx}, a simpler and more practical \gls{VB} scheme.

Unlike~\cite{Kinoshita_tight_2022}, in the current stage of our investigations, we apply \gls{MS-VBx} only at inference time. 
However, the training and inference mismatch can be compensated using the \gls{PLDA} model, which transforms the speaker embeddings for \gls{VBx}. Investigating tighter integration through joint training of \gls{EEND-VC} with \gls{MS-VBx} will be part of our future works.

\section{Experiments}
We evaluate the effectiveness of the proposed method for the CALLHOME~\cite{CALLHOME}, DIHARD II~\cite{DIHARDII}, and DIHARD III~\cite{DIHARDIII} (full set) datasets.

\subsection{Settings}
\textbf{Data: }
The training data comprised 5.5k hours of simulated mixtures created as in~\cite{Fujita_IS2019}. It uses Switchboard-2 (Phase I \& I\hspace{-.1em}I \& I\hspace{-.1em}I\hspace{-.1em}I), Switchboard Cellular (Part 1 \& 2), and the NIST Speaker Recognition Evaluation (2004 \& 2005 \& 2006 \& 2008) with noise from the MUSAN corpus~\cite{MUSAN} and simulated room impulse responses~\cite{Ko_2017}.
The mixtures were of up to 7 speakers, with an average silence duration between utterances of the same speakers of 2 sec ($\beta=2$).

For the CALLHOME dataset, we used the dev/test set definition of prior works~\cite{Horiguchi2020_EDA_EEND}, which amounts to 249 and 250 sessions for adaptation and test, respectively.

For each task, we used the development set for adaptation and hyper-parameter tuning and reported results on the test set.
We report results using estimated \gls{SAD}, corresponding to track 2 of DIHARD II and III challenges. Note that our system directly estimates the speech activity using \gls{EEND-VC} without using any external \gls{SAD}.

\textbf{EEND-VC configuration:}
We used a \gls{EEND-VC} model similar to that in~\cite{WavLM}. We base our implementation on the publicly available code.\footnote{\url{https://github.com/nttcslab-sp/EEND-vector-clustering}} It consists of six-stacked Transformer encoder blocks with eight attention heads. 
We used the pre-trained WavLM-large to obtain the input speech features to the \gls{EEND-VC} model as in~\cite{WavLM}. The input feature dimension was 1024, and the output dimension for each attention block was 256. To produce the final output, the encoder's output is projected with a linear layer into $C = 3$ output streams, where each output consists of the frame-by-frame speaker activity binary decisions and the speaker embedding of 256 dimensions.

We trained the \gls{EEND-VC} model using chunks of 15 sec and sub-chunks (or subsequences in~\cite{Horiguchi2021_EEND_offline_GLA}) of 5 sec, using a similar training scheme as~\cite{Horiguchi2021_EEND_offline_GLA}. 
We trained the model for 70 epochs with a batch size of 2048 and averaged the model over the last five epochs. We used the Adam optimizer with the learning rate scheduler introduced in~\cite{Transformer_Vaswani} with 25000 warm-up steps.

For adaptation, we retrained the averaged model on the adaptation set for three epochs. We fixed the WavLM parameters during training but retrained them during adaptation with a learning rate of $10^{-5}$ and a batch size of one. 
We performed adaptation using chunks of 30 sec and sub-chunks of 5 sec. 
At test time, we reduced the sub-chunk length to $1.5$ sec.
We set the threshold $\tau$ to detect the silent speaker at 0.05.

\textbf{Clustering parameters:}
We compare the proposed \gls{MS-VBx} with \gls{cAHC}~\cite{EEND-vector-clustering_Interspeech2021,WavLM}.
We normalized the embedding with the L2 norm, thus the maximum distance between two embeddings is $2$.
We reduced the embedding dimension to $32$ with a \gls{LDA} model trained on the adaptation set.
For \gls{cAHC}, we used the Euclid distance to cluster the embeddings. 
We tuned the distance threshold for \gls{AHC} for values between 1 and 0.8 on the dev set. 

For \gls{MS-VBx}, we trained the \gls{PLDA} model on the speaker embeddings obtained by processing the adaptation data with \gls{EEND-VC} and applied the corresponding transformation to the embeddings. We used $F_A=0.4$, $F_B=17$ and set the loop probability, $P_{\text{loop}}$, at 0.8. 
Besides, we applied a median filter on the predicted diarization results with a window of 1.0 sec for CALLHOME and 0.28 sec for DIHARD II and III. 

\textbf{Baseline \gls{VC} system (SAD + VBx + OSD):} We compare our diarization with a \gls{VC} diarization system, which is based on conventional (i.e., single stream) \gls{VBx}~\cite{landini2022bayesian}. First, we used a \gls{SAD} suited for telephone speech and based on time-delay neural networks and statistical pooling\footnote{\url{http://kaldi-asr.org/models/m4}} for CALLHOME and a \gls{SAD} suited for wide-band data released in pyannote~\cite{bredin2020pyannote} for DIHARD II and III. Then, we applied \gls{VBx}~\cite{landini2022bayesian} and assigned the second nearest speaker to the segments detected as overlapping with \gls{OSD}~\cite{otterson2007efficient}.

\textbf{Evaluation metrics:} We evaluated results in terms of \gls{DER} accounting for the overlap regions with a collar value of 0.25 sec for CALLHOME and 0 sec for DIHARD II and III. 
We used an estimated number of speakers for all experiments, where the number of speakers was obtained as in~\cite{EEND-vector-clustering_Interspeech2021} for \gls{cAHC} and as the number of remaining Gaussians with \gls{VBx}.
We also evaluated the speaker counting performance in terms of mean error (ME), $ME = \frac{1}{R} \sum_{r=1}^R | C_r - \widehat{C}_r |$,
where $C_r$ and $\widehat{C}_r$ are the actual and estimated number of speakers in recording $r$, and $R$ is the total number of recordings in the test set.

\subsection{Results}
\begin{table}[tb]
%\centering
\caption{DERs (\%) for CALLHOME (CH), DIHARD II (DH II), and DIHARD III (DH III). The numbers in parenthesis indicate the speaker counting performance in terms of ME.}% (lower the better) and $ACC$ (higher the better).}
\vspace{-2mm}
\label{tab:results}
\scalebox{0.9}[0.9]{
\begin{tabular}{l@{\hspace{0.2cm}} l@{\hspace{0.2cm}}c@{\hspace{0.2cm}}c@{\hspace{0.2cm}}c}
\toprule
\multicolumn{2}{c}{System} & CH & DH II & DH III \\ %& DER ACC DER & DER \\ 
\midrule
1&\gls{SAD} + \gls{VBx} + \gls{OSD} & 13.6 & 26.8 & 20.5 \\
2&\gls{SAD} + VBx + resegm.~\cite{bredin21_interspeech} & - & 26.3$^{\dag}$ & 19.3 \\
3 & EDA-TS-VAD \cite{wang2022target} & 11.2 & - & - \\
4 & USTC-NELSLIP DH III~\cite{wangustc_DIHARDIII} & - & - & 16.8 \\
5 &EEND-GLA~\cite{Horiguchi2021_EEND_offline_GLA} & 11.8 & 28.3 & 19.5 \\
6 &\gls{EEND-VC} w/ WavLM~\cite{WavLM} & 10.4 & - & -\\
\midrule
7 &\gls{EEND-VC} (cAHC)& 	11.1 (1.2) & 28.2 (3.2) & 19.3 (1.3)\\
8 &+ \gls{MS-VBx} (Proposed) &	10.4 (0.6) & 26.4 (1.1) & 18.2 (0.7) \\
\bottomrule
\end{tabular}}
\\
$^{\dag}$\footnotesize{results taken from~\cite{Horiguchi_23_online_EEND_GLA}. }
\vspace{-6mm}
\end{table}

Table \ref{tab:results} compares the \gls{DER} for various diarization systems and our proposed \gls{EEND-VC} with \gls{MS-VBx} on CALLHOME, DIHARD II and III datasets.
The upper part of the table shows the performance of competitive \gls{VC}-based systems (systems 1 and 2), \gls{TS-VAD}-based systems (systems 3 and 4), and \gls{EEND-VC}-based systems (systems 5 and 6). To the best of our knowledge, these systems represent the state-of-the-art for these tasks. System 7 consists of our baseline \gls{EEND-VC} baseline system using \gls{cAHC} for clustering. System 8 is the same system using the proposed \gls{MS-VBx}.

Our baseline \gls{EEND-VC} (system 7) reproduces the WavLM-based system\cite{WavLM} (system 6). It performs slightly worse on the CALLHOME dataset (i.e., \gls{DER} of 11.1 vs. 10.4 in~\cite{WavLM}), probably because of the difference in the simulated training data. However, it is still a relatively strong baseline, outperforming most prior works on the CALLHOME and DIHARD III datasets. The DIHARD III top system (system 4) performs significantly better, but it is a much more complex system, which combines several diarization approaches\cite{wangustc_DIHARDIII}.

By comparing systems 7 and 8, we confirm that using the proposed \gls{MS-VBx} reduces \gls{DER} compared to \gls{cAHC} for all three datasets. 
It also significantly reduces speaker counting errors by about 50\%.
\gls{EEND-VC} using the proposed \gls{MS-VBx} achieves similar or superior performance than most prior diarization systems on these tasks. 
These results demonstrate the potential of the proposed \gls{MS-VBx} for speaker diarization. 

\section{Conclusion}

In this paper, we have introduced \gls{MS-VBx}, which is an extension of the \gls{VBx} algorithm to perform clustering on the multi-stream embeddings generated by recent \gls{EEND-VC} diarization systems. We have demonstrated the potential of \gls{MS-VBx} on three popular datasets.
In future works, we plan to investigate joint-training of \gls{EEND-VC} system with the \gls{MS-VBx} clustering. We will also test the proposed \gls{MS-VBx} with other \gls{EEND-VC} frameworks such as \gls{EEND-GLA}\cite{Horiguchi2021_EEND_offline_GLA}.

\bibliographystyle{IEEEtran}
\bibliography{ref}

\begin{thebibliography}{10}
\providecommand{\url}[1]{#1}
\csname url@samestyle\endcsname
\providecommand{\newblock}{\relax}
\providecommand{\bibinfo}[2]{#2}
\providecommand{\BIBentrySTDinterwordspacing}{\spaceskip=0pt\relax}
\providecommand{\BIBentryALTinterwordstretchfactor}{4}
\providecommand{\BIBentryALTinterwordspacing}{\spaceskip=\fontdimen2\font plus
\BIBentryALTinterwordstretchfactor\fontdimen3\font minus
  \fontdimen4\font\relax}
\providecommand{\BIBforeignlanguage}[2]{{%
\expandafter\ifx\csname l@#1\endcsname\relax
\typeout{** WARNING: IEEEtran.bst: No hyphenation pattern has been}%
\typeout{** loaded for the language `#1'. Using the pattern for}%
\typeout{** the default language instead.}%
\else
\language=\csname l@#1\endcsname
\fi
#2}}
\providecommand{\BIBdecl}{\relax}
\BIBdecl

\bibitem{DIHARD_JHU}
G.~Sell, D.~Snyder, A.~McCree, D.~Garcia-Romero, J.~Villalba, M.~Maciejewski,
  V.~Manohar, N.~Dehak, D.~Povey, S.~Watanabe, and S.~Khudanpur, ``Diarization
  is hard: Some experiences and lessons learned for the {JHU} team in the
  inaugural {DIHARD} challenge,'' in \emph{Proc. Interspeech}, 2018, pp.
  2808--2812.

\bibitem{Fujita_IS2019}
Y.~Fujita, N.~Kanda, S.~Horiguchi, K.~Nagamatsu, and S.~Watanabe, ``End-to-end
  neural speaker diarization with permutation-free objectives,'' in \emph{Proc.
  Interspeech}, 2019, pp. 4300--4304.

\bibitem{Horiguchi2022_EDA_EEND}
S.~Horiguchi, Y.~Fujita, S.~Watanabe, Y.~Xue, and P.~García, ``Encoder-decoder
  based attractors for end-to-end neural diarization,'' \emph{IEEE/ACM
  Transactions on Audio, Speech, and Language Processing}, vol.~30, pp.
  1493--1507, 2022.

\bibitem{Medennikov2020}
I.~Medennikov, M.~Korenevsky, T.~Prisyach, Y.~Khokhlov, M.~Korenevskaya,
  I.~Sorokin, T.~Timofeeva, A.~Mitrofanov, A.~Andrusenko, I.~Podluzhny,
  A.~Laptev, and A.~Romanenko, ``Target-speaker voice activity detection: {A}
  novel approach for multi-speaker diarization in a dinner party scenario,'' in
  \emph{Proc. Interspeech}, 2020, pp. 274--278.

\bibitem{EEND-vector-clustering_ICASSP2021}
K.~Kinoshita, M.~Delcroix, and N.~Tawara, ``Integrating end-to-end neural and
  clustering-based diarization: Getting the best of both worlds,'' in
  \emph{Proc. IEEE International Conference on Acoustics, Speech and Signal
  Processing (ICASSP)}, 2021, pp. 7198--7202.

\bibitem{Horiguchi2021_EEND_offline_GLA}
S.~Horiguchi, S.~Watanabe, P.~García, Y.~Xue, Y.~Takashima, and Y.~Kawaguchi,
  ``Towards neural diarization for unlimited numbers of speakers using global
  and local attractors,'' in \emph{Proc. IEEE Automatic Speech Recognition and
  Understanding Workshop (ASRU)}, 2021, pp. 98--105.

\bibitem{WavLM}
S.~Chen, C.~Wang, Z.~Chen, Y.~Wu, S.~Liu, Z.~Chen, J.~Li, N.~Kanda,
  T.~Yoshioka, X.~Xiao, J.~Wu, L.~Zhou, S.~Ren, Y.~Qian, Y.~Qian, J.~Wu,
  M.~Zeng, X.~Yu, and F.~Wei, ``{WavLM}: Large-scale self-supervised
  pre-training for full stack speech processing,'' \emph{IEEE Journal of
  Selected Topics in Signal Processing}, vol.~16, no.~6, pp. 1505--1518, 2022.

\bibitem{diez19_interspeech}
M.~Diez, L.~Burget, S.~Wang, J.~Rohdin, and J.~Černocký, ``Bayesian {HMM}
  based x-vector clustering for speaker diarization,'' in \emph{Proc.
  Interspeech}, 2019, pp. 346--350.

\bibitem{landini2022bayesian}
F.~Landini, J.~Profant, M.~Diez, and L.~Burget, ``Bayesian {HMM} clustering of
  x-vector sequences ({VBx}) in speaker diarization: {T}heory, implementation
  and analysis on standard tasks,'' \emph{Computer Speech \& Language},
  vol.~71, p. 101254, 2022.

\bibitem{landini2020but}
F.~Landini, S.~Wang, M.~Diez, L.~Burget, P.~Mat{\v{e}}jka,
  K.~{\v{Z}}mol{\'\i}kov{\'a}, L.~Mo{\v{s}}ner, A.~Silnova, O.~Plchot,
  O.~Novotn{\'y} \emph{et~al.}, ``{BUT} system for the second {DIHARD} speech
  diarization challenge,'' in \emph{Proc. IEEE International Conference on
  Acoustics, Speech and Signal Processing (ICASSP)}, 2020, pp. 6529--6533.

\bibitem{wangustc_DIHARDIII}
Y.~Wang, M.~He, S.~Niu, L.~Sun, T.~Gao, X.~Fang, J.~Pan, J.~Du, and C.-H. Lee,
  ``{USTC-NELSLIP} system description for {DIHARD-III} challenge,'' in
  \emph{Proc. The Third DIHARD Speech Diarization Challenge Workshop}, 2021.

\bibitem{yu2022summary}
F.~Yu, S.~Zhang, P.~Guo, Y.~Fu, Z.~Du, S.~Zheng, W.~Huang, L.~Xie, Z.-H. Tan,
  D.~Wang \emph{et~al.}, ``Summary on the {ICASSP} 2022 multi-channel
  multi-party meeting transcription grand challenge,'' in \emph{Proc. IEEE
  International Conference on Acoustics, Speech and Signal Processing
  (ICASSP)}, 2022, pp. 9156--9160.

\bibitem{Fujita_ASRU2019}
Y.~Fujita, N.~Kanda, S.~Horiguchi, Y.~Xue, K.~Nagamatsu, and S.~Watanabe,
  ``End-to-end neural speaker diarization with self-attention,'' in \emph{Proc.
  IEEE Workshop on Automatic Speech Recognition \& Understanding (ASRU)}, 2019,
  pp. 296--303.

\bibitem{EEND-vector-clustering_Interspeech2021}
K.~Kinoshita, M.~Delcroix, and N.~Tawara, ``Advances in integration of
  end-to-end neural and clustering-based diarization for real conversational
  speech,'' in \emph{Proc. Interspeech}, 2021, pp. 3565--3569.

\bibitem{COP-kmeans}
K.~Wagstaff, C.~Cardie, S.~Rogers, and S.~S.~Schroedl, ``Constrained k-means
  clustering with background knowledge,'' in \emph{Proc. 18th International
  Conference on Machine Learning (ICML)}, 2001.

\bibitem{Yang2013_CLC_Kmeans}
Y.~Yang, T.~Rutayisire, C.~Lin, T.~Li, and F.~Teng, ``An improved {Cop-Kmeans}
  clustering for solving constraint violation based on {MapReduce} framework,''
  \emph{Fundam. Inf.}, vol. 126, no.~4, p. 301–318, 2013.

\bibitem{constrained_AHC}
I.~Davidson and S.~S. Ravi, ``Using instance-level constraints in agglomerative
  hierarchical clustering: {T}heoretical and empirical results,'' \emph{Data
  Mining and Knowledge Discovery}, vol.~77, no.~18, pp. 257--282, 2009.

\bibitem{Diez_taslp_2020}
M.~Diez, L.~Burget, F.~Landini, and J.~Černocký, ``Analysis of speaker
  diarization based on bayesian {HMM} with eigenvoice priors,'' \emph{IEEE/ACM
  Transactions on Audio, Speech, and Language Processing}, vol.~28, pp.
  355--368, 2020.

\bibitem{Kinoshita_tight_2022}
K.~Kinoshita, M.~Delcroix, and T.~Iwata, ``Tight integration of neural- and
  clustering-based diarization through deep unfolding of infinite {Gaussian}
  mixture model,'' in \emph{Proc. IEEE International Conference on Acoustics,
  Speech and Signal Processing (ICASSP)}, 2022, pp. 8382--8386.

\bibitem{CALLHOME}
M.~Przybocki and A.~Martin, \emph{2000 {NIST} Speaker Recognition Evaluation
  {(LDC2001S97)}}.\hskip 1em plus 0.5em minus 0.4em\relax Philadelphia, New
  Jersey: Linguistic Data Consortium, 2001.

\bibitem{DIHARDII}
N.~Ryant, K.~Church, C.~Cieri, A.~Cristia, J.~Du, S.~Ganapathy, and
  M.~Liberman, ``The second {DIHARD} diarization challenge: Dataset, task, and
  baselines,'' in \emph{Proc. Interspeech}, 2019, pp. 978--982.

\bibitem{DIHARDIII}
N.~Ryant, P.~Singh, V.~Krishnamohan, R.~Varma, K.~Church, C.~Cieri, J.~Du,
  S.~Ganapathy, and M.~Liberman, ``The third {DIHARD} diarization challenge,''
  in \emph{Proc. Interspeech}, 2021, pp. 3570--3574.

\bibitem{MUSAN}
D.~Snyder, G.~Chen, and D.~Povey, ``{MUSAN}: A music, speech, and noise
  corpus,'' 2015, arXiv:1510.08484.

\bibitem{Ko_2017}
T.~Ko, V.~Peddinti, D.~Povey, M.~L. Seltzer, and S.~Khudanpur, ``A study on
  data augmentation of reverberant speech for robust speech recognition,'' in
  \emph{Proc. IEEE International Conference on Acoustics, Speech and Signal
  Processing (ICASSP)}, 2017, pp. 5220--–5224.

\bibitem{Horiguchi2020_EDA_EEND}
S.~Horiguchi, Y.~Fujita, S.~Watanabe, Y.~Xue, and K.~Nagamatsu, ``End-to-end
  speaker diarization for an unknown number of speakers with encoder-decoder
  based attractors,'' in \emph{Proc. Interspeech}, 2020, pp. 269--273.

\bibitem{Transformer_Vaswani}
A.~Vaswani, N.~Shazeer, N.~Parmar, J.~Uszkoreit, L.~Jones, A.~N. Gomez,
  L.~Kaiser, and I.~Polosukhin, ``Attention is all you need,'' in \emph{Proc.
  The Thirty-first Annual Conference on Neural Information Processing Systems
  (NIPS)}, 2017, pp. 5998--–6008.

\bibitem{bredin2020pyannote}
H.~Bredin, R.~Yin, J.~M. Coria, G.~Gelly, P.~Korshunov, M.~Lavechin, D.~Fustes,
  H.~Titeux, W.~Bouaziz, and M.-P. Gill, ``Pyannote. audio: neural building
  blocks for speaker diarization,'' in \emph{Proc. IEEE International
  Conference on Acoustics, Speech and Signal Processing (ICASSP)}, 2020, pp.
  7124--7128.

\bibitem{otterson2007efficient}
S.~Otterson and M.~Ostendorf, ``Efficient use of overlap information in speaker
  diarization,'' in \emph{Proc. IEEE Workshop on Automatic Speech Recognition
  \& Understanding (ASRU)}, 2007, pp. 683--686.

\bibitem{bredin21_interspeech}
H.~Bredin and A.~Laurent, ``End-to-end speaker segmentation for overlap-aware
  resegmentation,'' in \emph{Proc. Interspeech}, 2021, pp. 3111--3115.

\bibitem{wang2022target}
D.~Wang, X.~Xiao, N.~Kanda, T.~Yoshioka, and J.~Wu, ``Target speaker voice
  activity detection with transformers and its integration with end-to-end
  neural diarization,'' \emph{arXiv preprint arXiv:2208.13085}, 2022.

\bibitem{Horiguchi_23_online_EEND_GLA}
S.~Horiguchi, S.~Watanabe, P.~García, Y.~Takashima, and Y.~Kawaguchi, ``Online
  neural diarization of unlimited numbers of speakers using global and local
  attractors,'' \emph{IEEE/ACM Transactions on Audio, Speech, and Language
  Processing}, vol.~31, pp. 706--720, 2023.

\end{thebibliography}

\end{document}